\documentclass[reprint,superscriptaddress,amsmath,amssymb,aps,prc,nofootinbib]{revtex4-1}
\usepackage{amsmath,amssymb}
\usepackage{bm}
\usepackage{graphicx}
\usepackage{dcolumn}
\usepackage{color}
\usepackage[colorinlistoftodos]{todonotes}
\usepackage[colorlinks,citecolor=blue]{hyperref}

\newcommand{\be}{\begin{equation}}
\newcommand{\ee}{\end{equation}}
\newcommand{\bea}{\begin{eqnarray}}
\newcommand{\eea}{\end{eqnarray}}

\newcommand{\mbss}[1]{_{\mbox{\scriptsize #1}}}

\newcommand{\vphu}{\vphantom{*}}

\newcommand{\ve}{\varepsilon}

\begin{document}

\title{Self-consistent description of high-spin states in doubly magic
  $^{208}$Pb}
\author{N. Lyutorovich}
\affiliation{St. Petersburg State University, St. Petersburg, 199034,
  Russia}
\author{V. Tselyaev}
\affiliation{St. Petersburg State University, St. Petersburg, 199034,
  Russia}
\author{J. Speth}
\affiliation{Institut f\"ur Kernphysik,
  Forschungszentrum J\"ulich, D-52425 J\"ulich, Germany}
\author{G. Martinez-Pinedo} \affiliation{GSI Helmholtzzentrum f\"ur
  Schwerionenforschung, Planckstra{\ss}e~1, 64291 Darmstadt, Germany}
\affiliation{Institut f{\"u}r Kernphysik (Theoriezentrum), Fachbereich
  Physik, Technische Universit{\"a}t Darmstadt,
  Schlossgartenstra{\ss}e 2, 64298 Darmstadt, Germany}
\affiliation{Helmholtz Forschungsakademie Hessen f\"ur FAIR, GSI
  Helmholtzzentrum f\"ur Schwerionenforschung, Planckstra{\ss}e~1,
  64291 Darmstadt, Germany}
\author{K. Langanke}
\affiliation{GSI
  Helmholtzzentrum f\"ur Schwerionenforschung, Planckstra{\ss}e~1,
  64291 Darmstadt, Germany}
\affiliation{Institut f{\"u}r Kernphysik
  (Theoriezentrum), Fachbereich Physik, Technische Universit{\"a}t
  Darmstadt, Schlossgartenstra{\ss}e 2, 64298 Darmstadt, Germany}
\author{P.-G. Reinhard}
\affiliation{Institut f\"ur Theoretische
  Physik II, Universit\"at Erlangen-N\"urnberg, D-91058 Erlangen,
  Germany}

\date{\today}

\begin{abstract}
We analyze recent data on a long series of high-spin states in
$^{208}$Pb with a self-consistent phonon-coupling model for nuclear
excitations based on the Skyrme functionals. The model is the
renormalized time-blocking approximation (RenTBA) which takes the
coherent one-particle-one-hole (1p1h) states of the random-phase
approximation (RPA) as starting point and develops from that more
complex configurations beyond RPA. To the best of our knowledge, this
is the first investigation of high spin states in $^{208}$Pb using
self-consistent nuclear models. The interesting point here is that
complex configurations are compulsory to describe the upper end of the
long spin series at all.  The data thus provide an ideal testing
ground for phonon-coupling models as they give direct access to
complex configurations. We find that standard Skyrme functionals which
perform well in ground state properties and giant resonance
excitations deliver at once an agreeable description of these high
spin states.
\end{abstract}

\maketitle

\section{Introduction}

High-spin states in nuclei show a rich variety of phenomena as, e.g.,
deformation alignment, back-bending, pairing breakdown, Coriolis
anti-pairing, having thus attracted much attention in the past, for
reviews see \cite{VoigtRMP83,SaladinBook91,Ward2001}. High-spin states
in the doubly magic $^{208}$Pb are special in that the large proton
and neutron shell gaps inhibit deformation and pairing. A description
in terms of an expansion into one-particle-one-hole (1p1h) states in a
space including the complete neutron and proton shells $1\hbar\omega$
above and below the Fermi surface is limited to angular momentum
$I\leq 14$ because the angular momenta in these shells cannot supply
more. The only way to couple to higher angular momenta are complex
configurations. This renders high-spin states in $^{208}$Pb to be a
unique laboratory to study complex configurations without
1p1h-background. The topic was taken up in a recent
paper~\cite{Broda_2017} which presents new experimental data up to
spin $I=30$ together with a theoretical analysis within large-scale
shell model calculations which are particularly suited to deal with
complex configurations. These shell model calculations consider one
proton and one neutron shell below and another proton and neutron
shell above the Fermi surface as valence space, tune the
single-particle (s.p.)  energies to the experimental spectra of the
neighboring odd nuclei, and use a microscopic two-body interaction. It
describes the experimental findings over the whole range of spins and
it allows to get insight into the structure of the states: States up
to $I^{\pi}=14^-$ are described by 1p1h configurations while spins up
to $I=26$ are dominated by 2p2h configurations and higher by 3p3h
configurations.

The aim of this paper is to see to what extent self-consistent models
can describe the measured series of high-spin states in $^{208}$Pb.
Self-consistent mean-field models on the basis of effective
energy-density functionals (EDF) manage to describe a wide range of
nuclear-structure properties and excitations, for reviews see
\cite{Bender_2003,Vretenar_2005,Erler_2011}.  Excited states are
usually described by the random-phase approximation (RPA) employing
consistently the same EDF as was used to prepare the ground state. The
RPA states are coherent superpositions of 1p1h configurations.  One way
to include more complex configurations is either to consider 2p2h
configurations explicitly \cite{Drozdz_1990} or to use phonon coupling
models where the 1p1h bases is coupled to nuclear ``phonons'' which
are themselves the more collective from the RPA states
\cite{Soloviev_1992,Col01a,Kamerdzhiev_2004}.  One of the major
successes is that phonon-coupling models provide a pertinent
description of the spreading widths and consequently deliver realistic
strength distributions for nuclear giant resonances, see
e.g. \cite{Lyutorovich_2012}. Still, the gross features of the
resonances particularly their average position are determined by RPA in
connection with proper choice of the EDF \cite{Kluepfel_2009}.

However, there are also states beyond reach of mere RPA.  An example
is the low-lying two-phonon states
\cite{Bertulani_1999,Brown_2000,Litvinova_2010,Litvinova_2013}.
Studies of such states beyond RPA within a self-consistent description
are so far scarce, although they provide an important further testing
ground for nuclear EDFs.  As argued above, high-spin states in
$^{208}$Pb are another welcome test case where we need to look beyond
RPA.  We will do this using the recently developed renormalized
time-blocking approximation (RenTBA) \cite{Tselyaev_2018} which is a
phonon-coupling model based on $\text{1p1h}\otimes\text{phonon}$
configurations where the phonons are self-consistently optimized
(renormalized) within the RenTBA loop.  RenTBA is a further developed
version of the time-blocking approximation (TBA)
\cite{Tselyaev_1989,Tselyaev_2007} taking into account the
single-particle continuum as described in
Refs.~\cite{Lyutorovich_2015,Lyutorovich_2016,Tselyaev_2016}.  The
upper end of the spin series of \cite{Broda_2017}, namely $I\geq 27$,
goes even beyond the $\text{1p1h}\otimes\text{phonon}$ space. For
these states we take into account $\text{2p2h}\otimes\text{phonon}$
configurations in an approximate RenTBA.

\section{Theoretical framework}
\label{sec:theor}

The calculations were made first in RenTBA \cite{Tselyaev_2018} where
the coupling between two $\text{1p1h}\otimes\text{phonon}$
configurations is realized through the RPA propagator entering the
Bethe-Salpeter equation for the response function taking into account
the single-particle continuum (see
\cite{Tselyaev_2018,Lyutorovich_2015,Lyutorovich_2016,Tselyaev_2016}).
RenTBA optimizes the phonons self-consistently within its own TBA
scheme (not taking simply phonons from RPA calculations). This allows
to achieve high quality with fewer basis phonon states which reduces
computations expense and the danger of double counting.

For the high-spin states in $^{208}$Pb with $14<I<27$, the
$\text{1p1h}\otimes\text{phonon}$ configurations lie energetically in
the nucleon continuum. Thus the coupling between them is mediated
by the continuum part of the RPA response function, for details see
Appendix \ref{app:highspin}. The continuum states couple only weakly
to higher configurations.
As a result, the uncoupled $\text{1p1h}\otimes\text{phonon}$ basis states
are already a good approximation to the final result.  We have
confirmed this assumption
by comparison with
the states from full RenTBA. Thus we introduce an approximate
RenTBA$_0$ where the coupling is neglected.
This becomes particularly simple for the energies which then read
\begin{equation}
 E_{ph\nu} = \varepsilon^{\vphu}_p-\varepsilon^{\vphu}_h+\omega_\nu
 \label{e1p1hphon}
\end{equation}
where the $\varepsilon^{\vphu}_\alpha$ are the s.p. energies from the ground
state EDF calculations and $\omega_\nu$ the eigen-energies of the
RenTBA phonons (which can be determined independently from the
high-spin states). The renormalization of the effective interaction is
mediated here exclusively through the phonon energy $\omega_\nu$.

The RenTBA$_0$ then allows us to proceed to the upper end of the
high-spin series with $I\geq 27$. This requires
$\text{2p2h}\otimes\text{phonon}$ configuration which cannot be
treated in RenTBA without substantial extension of the scheme. An
obvious generalization of RenTBA$_0$ is to consider uncoupled
2p2h$\otimes$phonon whose energy is then simply
\begin{equation}
 E_{pp'hh'\nu}
 =
 \varepsilon^{\vphu}_p+\varepsilon^{\vphu}_{p'}
 -\varepsilon^{\vphu}_h-\varepsilon^{\vphu}_{h'}+\omega_\nu
 \;.
\end{equation}
As RenTBA$_0$ worked so well for  $I< 27$, we expect
reliable predictions also for the higher spins.

The technical details of the calculations were the following.  Wave
functions and fields were represented on a spherical grid in
coordinate space.  The s.p. basis was discretized by imposing box
boundary condition with a box radius equal to 18~fm.  To check that
the results do not depend on the box size, some calculations were also
performed with the box radius = 27~fm.  The particle's energies
$\ve^{\vphu}_{p}$ were limited by the maximum value
$\ve^{\mbss{max}}_{p} = 100$ MeV.  The details of solving the
non-linear RenTBA equations are described in
Ref.~\cite{Tselyaev_2020}. To find the energies of the states, the
strength functions $S(E)$ were calculated with very small smearing
parameters ($\Delta = 1$ keV, some times 0.1 keV or even less) and the
energies were read off from the maxima in the $S(E)$.

The calculations employed Skyrme EDF with three parameter sets. The
set SV-bas is chosen because it provides a good description of
properties of many nuclei (see~\cite{Kluepfel_2008}). The choice of
the spin-spin part of the EDF leaves several options open
\cite{Pot10a}.  But spin-spin interactions may be important for the
excitations with unnatural parity which we also consider here.  We
choose the most extensive option to include all spin terms and fix
them by the assumption as if the EDF is derived from the expectation
value of the same zero-range momentum-dependent two-body Skyrme force
as the SV-bas \cite{Pot10a}, thus without any new parameters.
This straightforward extension, although
successful for describing odd nuclei \cite{Pot10a}, is found to be
somewhat insufficient for low lying magnetic modes, a problem which
was solved by slightly modifying spin terms and spin-orbit force of
the EDF \cite{Tselyaev_2019,Tselyaev_2020}.  We consider here two of
these modified functionals
used in our calculations, namely SV-bas$_{-0.44}$ and SKXm$_{-0.49}$.  The
first is a variant of SV-bas from \cite{Kluepfel_2008} and the
second of SKXm from \cite{Brown_1998}. The modification was designed
to describe the nuclear ground-state properties with approximately the
same accuracy as the original SKXm and SV-bas and to reproduce at the
same time the basic experimental characteristics of the $M1$
excitations in $^{208}$Pb within the RenTBA.  We have yet to see how
the performance of unnatural parity states with higher spin comes out.

\section{Results and discussion}

\begin{figure}[h]
\includegraphics[width=\linewidth]{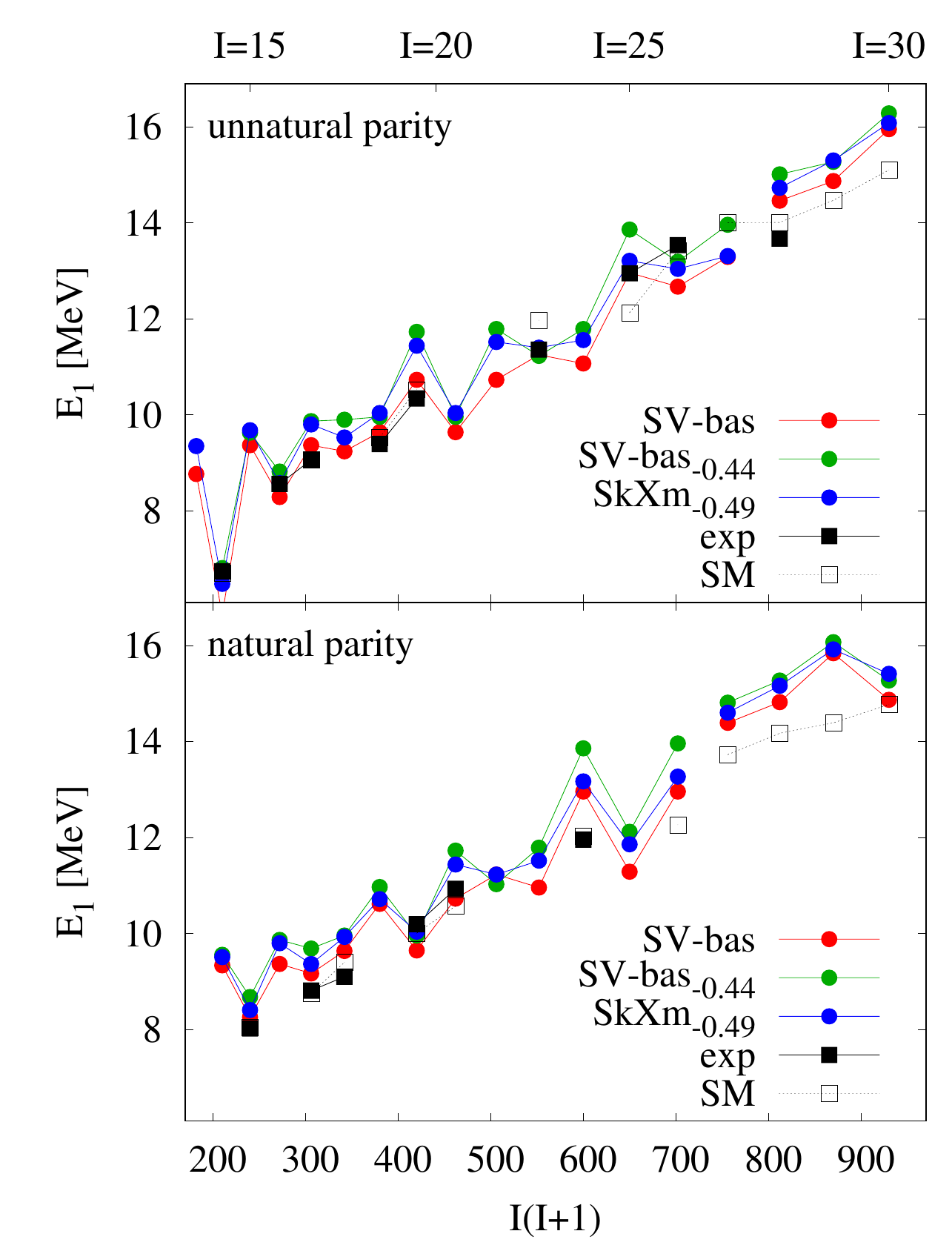}
\caption{\label{fig:Pb208-highspin} Energies $E_1$ of $^{208}$Pb yrast
  states plotted versus $I(I+1)$ for the spins $ 13 \leqslant I
  \leqslant 30$ computed with RenTBA for three different Skyrme
  parametrizations as indicated and compared with available experimental data
  as well as with shell model calculations (indicated
  by ``SM'') \cite{Broda_2017}.  The lower panel shows results
  for natural parity states and the upper panel for unnatural parity
  states. On the lower panel,  the lines are interrupted between $I=26$ and $I=27$ to
  indicate that the high $I$  states have a different structure
  (2p2h$\otimes$phonon). On the upper panel, the change appears  between $I=27$ and $I=28$.
  }
\end{figure}
\begin{figure}[h]
\includegraphics[width=\linewidth]{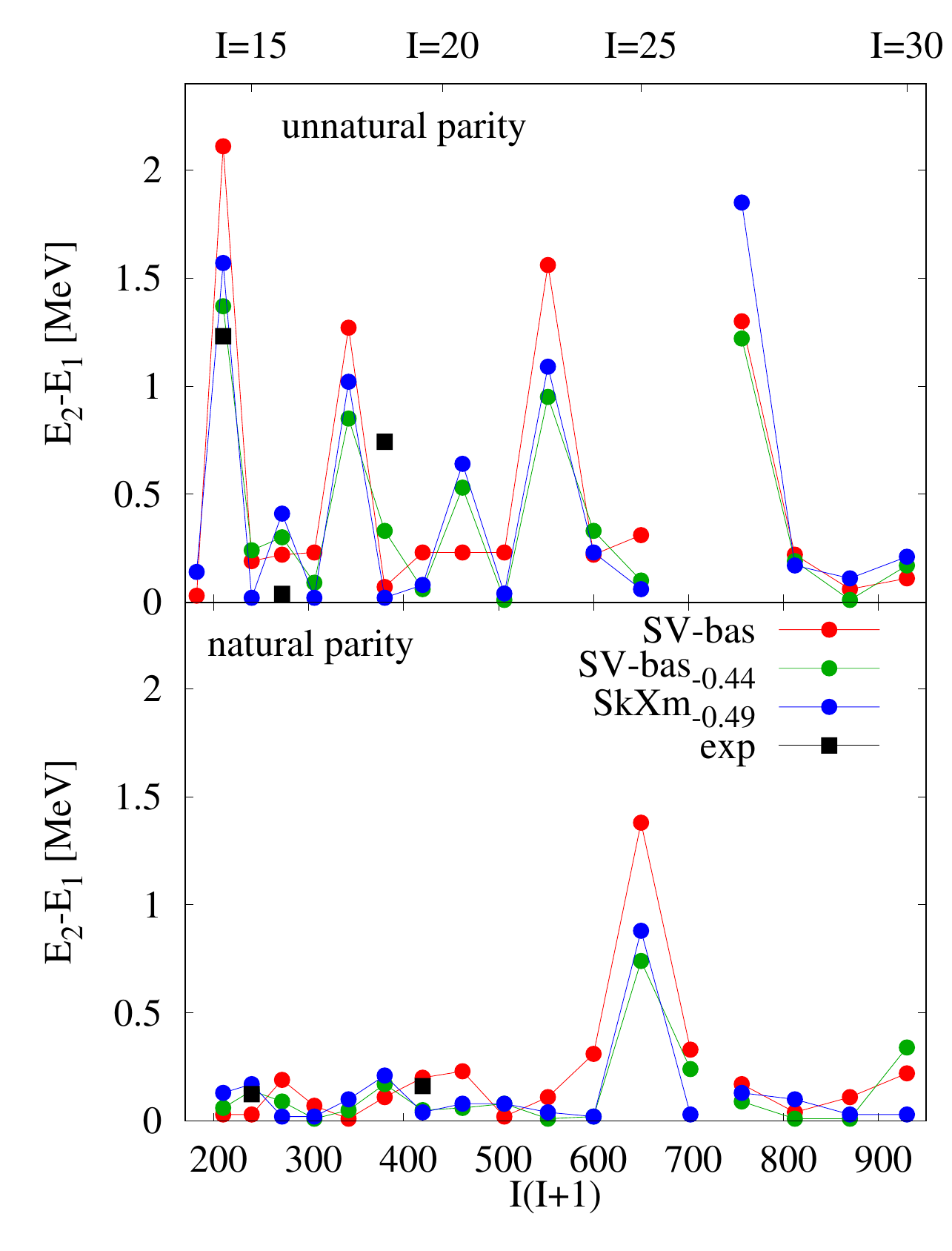}
\caption{\label{fig:Pb208-highspindiff} As figure
  \ref{fig:Pb208-highspin}, but for the energy difference $E_2-E_1$
  for the first two excited states in each angular momentum $I$
  channel.}
\end{figure}

The results for the $^{208}$Pb high-spin states are presented in
Figs. \ref{fig:Pb208-highspin} and \ref{fig:Pb208-highspindiff}, where
$I$ and $\pi$ denote spin and parity of a state, $E$ are experimental
and theoretical energies (in MeV). The experimental values
were taken from \cite{Broda_2017,NDS_2007}
and the shell
model values from \cite{Broda_2017}.
In some cases the spin and parity assignment of
levels do not coincide between Refs. \cite{Broda_2017} and
\cite{NDS_2007}, namely the 7.974, 8.027, and 9.061 MeV levels.  In
these cases, we preferred the identification \cite{Broda_2017}, since
it is the later one.  Our calculations confirm the correctness of this
choice.
The RenTBA energies calculated using the SV-bas, SV-bas$_{-0.44}$, and
SKXm$_{-0.49}$ parameter sets are shown in the figures by red, green,
and blue dots connected by lines of the same colors.  The experimental
energies are shown in filled black squares and the shell model results
in open black squares.
The experimental data are surprisingly well described by all
mean-field based models.
In particular, the trend with spin $I$ is well reproduced.
One can see, however, that within the three mean-field parametrizations
the SV-bas has a slight preference in spite of the fact that
the SV-bas$_{-0.44}$ and SKXm$_{-0.49}$ produce the improved spin residual interaction.
The reason is that the low-lying $1^+$ phonons and the respective
1p1h energies in the $^{208}$Pb which were used in the fit of the parameters of the
SV-bas$_{-0.44}$ and SKXm$_{-0.49}$ sets do not enter the configurations
relevant for the high-spin states, see Appendix \ref{app:complex}.
The shell model (open black squares) performs usually a bit better than
SV-bas. But there also cases where SV-bas wins.  Taken over all, both
models are competitive.

Finally, we remark that the general trend looks very much like a
rotational band, though the experimental trend in detail does not
always follow exactly a straight line ($E$ is approximately constant
in the ranges $I = 17-18$ and $26-28$). Nevertheless, taken over
  all both experimental and theoretical trends are similar to a
  rotational band. However, this band does not rely on a collective
  rotation in the sense of cranking \cite{Rin80aB}.  This is hindered
  by the spherical shape and the large shell gap of $^{208}$Pb.  It is
  not excluded to find a representation in terms of a rotating exotic
  deformation coupled to 1p1h states as was proposed, e.g., in
  \cite{Heusler_2020} for the high-spin states in the 6 MeV
  region. Our analysis of the microscopic structure shows that the
  high-spin states are of predominantly s.p. structure, composed from
  combinations of s.p. states and phonons with high spin. The
  rotational trend stems from a change in the angular momentum of one
  single nucleon at a time where the rotational part of the
  s.p. kinetic energy gives a large contribution to the s.p. energy,
  for quantitative details see appendix \ref{app:complex}.  In
  practice, our calculations show that the $13^-_1$ and $14^-_1$
  states have still mixed $\text{1p1h} +
  \text{1p1h}\otimes\text{phonon}$ structure while the states with
  spins $13\leqslant I \leqslant 26$ (except $26^-_2$) are
  predominantly 1p1h$\otimes$phonon configurations and the even larger
  $I$ need $\text{2p2h}\otimes\text{phonon}$ configurations.  The
  $26^-_2$ state is at the transition point. The energy of its
  $\text{1p1h}\otimes\text{phonon}$ configuration is 15.16~MeV which
  is higher than the 13.51~MeV of the
  $\text{2p2h}\otimes\text{phonon}$ state (values for SV-bas).  The
  latter is then the relevant configuration.

The data in \cite{Broda_2017} allow also to deduce a few values for
the second excited state. Drawing them together with the RenTBA
results looks very much like Fig.~\ref{fig:Pb208-highspin} for the
first excited state, a near linear trend with $I(I\!+\!1)$ and
acceptable agreement with theory. We do not show that here to avoid
doubling. Rather we take an amplifying glass and look in
Fig.~\ref{fig:Pb208-highspindiff} at the energy difference between second
and first excited states.  All three Skyrme parametrizations yield the
same trend, however with occasional visible quantitative differences.
The few experimental data basically agree with the theoretical values,
except for $I\!=\!18$. Unfortunately the data are still too sparse to
be conclusive.

\section{Conclusions}

In this paper, the energies of the high-spin yrast and near-yrast
states in $^{208}$Pb have been calculated within a fully
self-consistent approach beyond RPA taking into account 1p1h and
1p1h$\otimes$phonon configurations.  The approach is based on the
Skyrme energy-density functional (EDF) and on the recently developed
renormalized version of the time-blocking approximation (RenTBA)
\cite{Tselyaev_2018}.  The results are compared to the recent
experimental data and shell model results \cite{Broda_2017}.  The main
conclusions of our calculations are the following:
\begin{enumerate}
\item
The self-consistent approach based on the Skyrme EDF and the RenTBA
give reasonable agreement with the experimental energies
of the high-spin yrast states in $^{208}$Pb without refitting the
parameters of the Skyrme EDF previously determined from the large-scale
calculations of many nuclear properties.
\item
The yrast states with the total angular momenta $15 \leqslant I
\leqslant 26$ in $^{208}$Pb can be interpreted with high accuracy as
the pure 1p1h$\otimes$phonon configurations in our model,
those with even higher spins as 2p2h$\otimes$phonon configurations.
\item
The energies of the phonons used in these 1p1h$\otimes$phonon
configurations should be determined in the framework of the full-scale
RenTBA that implies the solution of the system of nonlinear equations
of this model.
\end{enumerate}

\begin{acknowledgements}
This work was supported by the Russian Foundation for Basic Research,
project number 21-52-12035, and the Deutsche Forschungsgemeinschaft,
contract RE 322/15-1.  This research was carried out using
computational resources provided by the Computer Center of
St. Petersburg State University.
\end{acknowledgements}

\appendix

\section{Details on the high-spin states}

\subsection{High spins and 1p1h or RPA states}
\label{app:highspin}

All excited states in $^{208}$Pb with spins $I \geqslant 15$ lie above the neutron
and proton separation energies which are, e.g., $S_n = 7.58$ and $S_p = 7.94$ MeV
for the SV-bas.
Therefore, discrete RPA (DRPA) becomes inappropriate
because the artificial quantizatation of continuum states
by DRPA falsifies energies and strengths.
\begin{table}[h]
\caption{\label{table:RPA}
DRPA results for the energies (in MeV) of the high-spin yrast states of
$^{208}$Pb in comparison with the experimental data \cite{Broda_2017}.
Here, $r_b$ denotes the box radius (in fm) used in the DRPA calculations.
}
\begin{ruledtabular}
\begin{tabular}{crcc}

               & \multicolumn{1}{c}{Exp.}    & \multicolumn{2}{c}{DRPA} \\
\cline{3-4}
               &         &$r_b=18$ & $r_b =27$ \\

\hline
$E(15^-_1)$    & 8.027   & 21.5    & 14.7  \\
$E(17^+_1)$    & 9.061   & 23.6    & 15.6  \\
$E(20^-_1)$    & 10.342  & 30.7    & 18.7   \\
$E(20^+_1)$    & 10.196  & 33.3    & 19.9   \\
\end{tabular}
\end{ruledtabular}
\end{table}
One signature of that is a strong dependence of these states on box
size as illustrated in Table~\ref{table:RPA} for two values of the box
radius $r_b$.
The continuum RPA (CRPA) correctly takes into account the single-particle continuum
for every $I^\pi$. However, the CRPA strength functions for the high-spin
electric or magnetic excitations in $^{208}$Pb have one
dominant and very broad peak whose centroid is in the region above 20 MeV
with a width of 15-25 MeV depending on the excitation.
These centroid energies stay way above the experimental energies
of the high-spin yrast states in $^{208}$Pb
that shows that mere RPA is unable to describe these data.

The continuum structure of 1p1h (or RPA) states has
consequences for their coupling to 1p1h$\otimes$phonon states and for
the coupling between the 1p1h$\otimes$phonon states.
The 1p1h$\otimes$phonon configurations considered in our study
and the matrix elements of the interaction between them and the 1p1h configurations
are formed predominantly by the discrete and quasidiscrete s.p. states
which are orthogonal or approximately orthogonal to the continuum states.
For this reason the couplings mentioned above are weak.
Nonetheless, the coupling between the 1p1h and 1p1h$\otimes$phonon states
is necessary to provide the latter with some direct multipole strength.
The resulting multipole strength function has very narrow peaks
with the small integral strength in the vicinity of the energies of
1p1h$\otimes$phonon states and must be calculated there with high energy resolution.

\subsection{The structure of phonons and complex states}
\label{app:complex}

Table \ref{table:phonons} indicates the structure of the phonons in
percentage 1p1h components whereby only sufficiently large components
are shown. With few exceptions, states with high spins come close to
pure 1p1h states while low-spin states are typically a superposition
of many different 1p1h components. The latter can be called collective
phonons while the high-spin states are non-collective phonons.

\begin{table}
\caption{\label{table:phonons} Composition of phonons up
the highest spin which can be represented by mere RPA states.
Only 1p1h components with sufficiently large contributions
are given.
}
\begin{ruledtabular}
\begin{tabular}{lcl}
$3^-_{  1}$ &=& $ \nu\, 2g_{ 9/2} \, 3p_{ 3/2}{}^{-1}\,22\%
           + \pi\, 1h_{ 9/2} \, 2d_{ 3/2}{}^{-1}\,20\% + \dots$ \\
$4^+_{  1}$ &=& $ \nu\, 2g_{ 9/2} \, 1i_{13/2}{}^{-1}\, 55\%
            + \pi\, 1h_{ 9/2} \, 1h_{11/2}{}^{-1}\, 14\%$ \\
$5^-_{  1}$ &=& $ \nu\, 2g_{ 9/2} \, 3p_{ 1/2}{}^{-1} \,81\%$ \\
$6^-_{  1}$ &=& $ \nu\, 2g_{ 9/2} \, 3p_{ 3/2}{}^{-1}\,80\%
            + \nu\, 2g_{ 9/2} \, 2f_{ 5/2}{}^{-1}\,20\%$ \\
$6^+_{  1}$ &=& $ \pi\, 1i_{13/2} \, 3s_{ 1/2}{}^{-1}\,42\%
            + \nu\, 2g_{ 9/2} \, 1i_{13/2}{}^{-1}\,25\%$ \\
$7^-_{  1}$ &=& $ \nu\, 2g_{ 9/2} \, 2f_{ 5/2}{}^{-1} \,98\%$ \\
$8^-_{  1}$ &=& $ \nu\, 1i_{11/2} \, 2f_{ 5/2}{}^{-1}  \,99\%$ \\
$8^+_{  1}$ &=& $ \nu\, 2g_{ 9/2} \, 1i_{13/2}{}^{-1} \,83\%$ \\
$10^-_{  1}$ &=& $ \pi\, 1i_{13/2} \, 1h_{11/2}{}^{-1} \,54\%
            + \nu\, 1j_{15/2} \, 1i_{13/2}{}^{-1}\,45\%$ \\
$10^+_{  1}$ &=& $ \nu\, 2g_{ 9/2} \, 1i_{13/2}{}^{-1}\, 98\% $ \\
$11^+_{  1}$ &=& $ \nu\, 2g_{ 9/2} \, 1i_{13/2}{}^{-1} \, 100\%$ \\
$12^-_{  1}$ &=& $ \nu\, 1j_{15/2} \, 1i_{13/2}{}^{-1}\, 62\%
             + \pi\, 1i_{13/2} \, 1h_{11/2}{}^{-1}\, 38\% $ \\
$12^+_{  1}$ &=& $ \nu\, 1i_{11/2} \, 1i_{13/2}{}^{-1} \, 100\%$ \\
$13^-_{  1}$ &=& $ \nu\, 1j_{15/2} \, 1i_{13/2}{}^{-1}\, 100\%$ \\
$13^+_{  1}$ &=& $ \nu\, 1j_{15/2} \, 1h_{11/2}{}^{-1}\, 88\%$ \\
$14^-_{  1}$ &=& $ \nu\, 1j_{15/2} \, 1i_{13/2}{}^{-1}\, 100\% $
\end{tabular}
\end{ruledtabular}
\end{table}

Table \ref{table:multiphonon} shows the energies and structures of the
high-spin excitations at the level of 2p2h, 1p1h$\otimes$phonon, and
two-phonon states
calculated with SKXm$_{-0.49}$. That makes it obvious that the states
with very high spin are composed from s.p. states with high spin via
phonons with high spins. It becomes also apparent that the states
acquire increasingly simple structures, the more so as the strength of
the residual interaction decreases with increasing spin.

\begin{table*}
\caption{\label{table:multiphonon} Energies and structures of 2p2h,
1p1h$\otimes$phonon, and two-phonon energies for high spins calculated
with SKXm$_{-0.49}$.
The energies were calculated for
the renormalized phonons.
}
\begin{ruledtabular}
\begin{tabular}{cccccccc}
$I^\pi_n$& Exp.  & \multicolumn{2}{c}{2p2h}
                                  &\multicolumn{2}{c}{1p1h$\otimes$phonon}
                                                  &\multicolumn{2}{c}{phonon$\otimes$phonon}\\
                  \cline{3-4}      \cline{5-6}      \cline{7-8}
         &       & $E$   & Configuration
                                  & $E$   & Configuration
                                                   & $E$  & Configuration \\
\hline
$15^-_1$ & 8.027 & 9.10  & $\nu\, 2g_{ 9/2} \, 1i_{13/2}{}^{-1}
                            \otimes \nu\, 2g_{ 9/2} \, 3p_{ 1/2}{}^{-1}$
                                  & 8.41  & $\nu\, 2g_{ 9/2} \, 3p_{ 1/2}{}^{-1} \otimes 10^+_{  1}$
                                                   & 7.89  & $10^+_{  1} \otimes  5^-_{  1}$  \\
$15^+_1$ & \footnotemark[1]
                 & 10.56 & $\nu\, 1i_{11/2} \, 2f_{ 5/2}{}^{-1}
                            \otimes \nu\, 2g_{ 9/2} \, 2f_{ 5/2}{}^{-1}$
                                  & 9.68  & $\nu\, 2g_{ 9/2} \, 1i_{13/2}{}^{-1} \otimes  4^+_{  1}$
                                                   & 8.83  & $12^-_{  1} \otimes  3^-_{  1}$ \\
$16^-_1$ & 8.562 \footnotemark[2]
                 & 9.10  & $\nu\, 2g_{ 9/2} \, 1i_{13/2}{}^{-1}
                            \otimes \nu\, 2g_{ 9/2} \, 3p_{ 1/2}{}^{-1}$
                                  & 8.58  & $\nu\, 2g_{ 9/2} \, 3p_{ 1/2}{}^{-1} \otimes 11^+_{  1}$
                                                   & 8.06  & $11^+_{  1} \otimes  5^-_{  1}$  \\
$16^+_1$ &       & 10.70  & $\nu\, 2g_{ 9/2} \, 1i_{13/2}{}^{-1}
                             \otimes \pi\, 1i_{13/2} \, 3s_{ 1/2}{}^{-1}$
                                  & 9.80  & $\nu\, 1j_{15/2} \, 1i_{13/2}{}^{-1} \otimes  3^-_{  1}$
                                                   & 8.87  & $13^-_{  1} \otimes  3^-_{  1}$  \\
$17^-_1$ & 8.8128 & 9.92  & $\nu\, 2g_{ 9/2} \, 1i_{13/2}{}^{-1}
                            \otimes \nu\, 2g_{ 9/2} \, 3p_{ 3/2}{}^{-1}$
                                  & 9.36  & $\nu\, 2g_{ 9/2} \, 2f_{ 5/2}{}^{-1} \otimes 10^+_{  1}$
                                                   & 8.92  & $11^+_{  1} \otimes  6^-_{  1}$  \\
$17^+_1$ & 9.061 \footnotemark[1]
                 & 11.67 & $\nu\, 1j_{15/2} \, 1i_{13/2}{}^{-1}
                            \otimes \nu\, 2g_{ 9/2} \, 3p_{ 3/2}{}^{-1}$
                                  & 9.80  & $\nu\, 1j_{15/2} \, 1i_{13/2}{}^{-1} \otimes  3^-_{  1}$
                                                   & 9.17  &  $14^-_{  1} \otimes  3^-_{  1}$ \\
$18^-_1$ &       & 10.06  & $\nu\, 2g_{ 9/2} \, 1i_{13/2}{}^{-1}
                             \otimes \nu\, 2g_{ 9/2} \, 2f_{ 5/2}{}^{-1}$
                                  & 9.54  & $\nu\, 2g_{ 9/2} \, 2f_{ 5/2}{}^{-1} \otimes 11^+_{  1}$
                                                   & 9.22  & $11^+_{  1} \otimes  7^-_{  1}$  \\
$18^+_1$ & 9.1030& 10.70  & $\nu\, 2g_{ 9/2} \, 1i_{13/2}{}^{-1}
                             \otimes \pi\, 1i_{13/2} \, 3s_{ 1/2}{}^{-1}$
                                  & 9.92  & $\nu\, 2g_{ 9/2} \, 3p_{ 1/2}{}^{-1} \otimes 13^-_{  1}$
                                                   & 9.35  & $10^+_{  1} \otimes  8^+_{  1}$  \\
$19^-_1$ & 9.394 & 11.24 & $\nu\, 1i_{11/2} \, 1i_{13/2}{}^{-1}
                             \otimes \nu\, 2g_{ 9/2} \, 2f_{ 5/2}{}^{-1}$
                                  & 10.71 & $\nu\, 1i_{11/2} \, 2f_{ 5/2}{}^{-1} \otimes 11^+_{  1}$
                                                   & 10.30 & $11^+_{  1} \otimes  8^-_{  1}$ \\
$19^+_1$ & 9.394 & 10.73  & $\nu\, 2g_{ 9/2} \, 1i_{13/2}{}^{-1}
                             \otimes \nu\, 2g_{ 9/2} \, 1i_{13/2}{}^{-1}$
                                  & 10.04 & $\nu\, 2g_{ 9/2} \, 1i_{13/2}{}^{-1} \otimes  8^+_{  1}$
                                                   & 9.35  & $10^+_{  1} \otimes 10^+_{  1}$  \\
$20^-_1$ &10.3419 & 12.39  & $\pi\, 1i_{13/2} \, 1h_{11/2}{}^{-1}
                             \otimes \nu\, 2g_{ 9/2} \, 1i_{13/2}{}^{-1}$
                                  & 11.44 & $\nu\, 2g_{ 9/2} \, 1i_{13/2}{}^{-1} \otimes 10^-_{  1}$
                                                   & 10.75 & $10^-_{  1} \otimes 10^+_{  1}$  \\
$20^+_1$ &10.1959& 10.73  & $\nu\, 2g_{ 9/2} \, 1i_{13/2}{}^{-1}
                             \otimes \nu\, 2g_{ 9/2} \, 1i_{13/2}{}^{-1}$
                                  & 10.04 & $\nu\, 2g_{ 9/2} \, 1i_{13/2}{}^{-1} \otimes 10^+_{  1}$
                                                    & 9.35 & $10^+_{  1} \otimes 10^+_{  1}$  \\
$21^-_1$ &10.9343& 12.39  & $\pi\, 1i_{13/2} \, 1h_{11/2}{}^{-1}
                             \otimes \nu\, 2g_{ 9/2} \, 1i_{13/2}{}^{-1}$
                                  & 11.44 & $\nu\, 2g_{ 9/2} \, 1i_{13/2}{}^{-1} \otimes 10^-_{  1}$
                                                    & 10.83 & $12^-_{  1} \otimes 10^+_{  1}$  \\
$21^+_1$ &       & 10.73 & $\nu\, 2g_{ 9/2} \, 1i_{13/2}{}^{-1}
                            \otimes \nu\, 2g_{ 9/2} \, 1i_{13/2}{}^{-1}$
                                  & 10.04 & $\nu\, 2g_{ 9/2} \, 1i_{13/2}{}^{-1} \otimes 10^+_{  1}$
                                                    & 9.52  & $11^+_{  1} \otimes 10^+_{  1}$ \\
$22^-_1$ &       & 12.39  & $\nu\, 2g_{ 9/2} \, 1i_{13/2}{}^{-1}
                             \otimes \pi\, 1i_{13/2} \, 1h_{11/2}{}^{-1}$
                                  & 11.52 & $\nu\, 2g_{ 9/2} \, 1i_{13/2}{}^{-1} \otimes 12^-_{  1}$
                                                    & 10.83 & $12^-_{  1} \otimes 10^+_{  1}$  \\
$22^+_1$ &       & 11.91  &$\nu\, 1i_{11/2} \, 1i_{13/2}{}^{-1}
                           \otimes \nu\, 2g_{ 9/2} \, 1i_{13/2}{}^{-1} $
                                  & 11.22 & $\nu\, 1i_{11/2} \, 1i_{13/2}{}^{-1} \otimes 10^+_{  1}$
                                                    & 10.78  & $11^+_{  2} \otimes 11^+_{  1}$ \\
$23^-_1$ &       & 12.39  & $\pi\, 1i_{13/2} \, 1h_{11/2}{}^{-1}
                             \otimes \nu\, 2g_{ 9/2} \, 1i_{13/2}{}^{-1}$
                                  & 11.52 & $\nu\, 2g_{ 9/2} \, 1i_{13/2}{}^{-1} \otimes 12^-_{  1}$
                                                    & 10.87  & $13^-_{  1} \otimes 10^+_{  1}$ \\
$23^+_1$ & 11.3609
                 & 11.92 & $\nu\, 2g_{ 9/2} \, 1i_{13/2}{}^{-1}
                            \otimes \nu\, 1i_{11/2} \, 1i_{13/2}{}^{-1}$
                                  & 11.40 & $\nu\, 1i_{11/2} \, 1i_{13/2}{}^{-1} \otimes 11^+_{  1}$
                                                    & 11.27 & $11^+_{  1} \otimes 12^+_{  1}$ \\
$24^-_1$ &       & 12.48 & $\nu\, 2g_{ 9/2} \, 1i_{13/2}{}^{-1}
                            \otimes \nu\, 1j_{15/2} \, 1i_{13/2}{}^{-1}$
                                  & 11.56 & $\nu\, 2g_{ 9/2} \, 1i_{13/2}{}^{-1} \otimes 13^-_{  1}$
                                                    & 11.04 & $11^+_{  1} \otimes 13^-_{  1}$ \\
$24^+_1$ &11.9582& 14.14 & $\nu\, 1j_{15/2} \, 1i_{13/2}{}^{-1}
                            \otimes \pi\, 1i_{13/2} \, 1h_{11/2}{}^{-1}$
                                  & 13.17 & $\pi\, 1i_{13/2} \, 1h_{11/2}{}^{-1} \otimes 12^-_{  1}$
                                                    & 12.30 & $12^-_{  1} \otimes 12^-_{  1}$ \\
$25^-_1$ &       & 12.48 & $\nu\, 1j_{15/2} \, 1i_{13/2}{}^{-1}
                            \otimes \nu\, 2g_{ 9/2} \, 1i_{13/2}{}^{-1}$
                                  & 11.86 & $\nu\, 2g_{ 9/2} \, 1i_{13/2}{}^{-1} \otimes 14^-_{  1}$
                                                    & 11.34  & $14^-_{  1} \otimes 11^+_{  1}$ \\
$25^+_1$ & 12.9493
                 & 14.14 & $\pi\, 1i_{13/2} \, 1h_{11/2}{}^{-1}
                            \otimes \nu\, 1j_{15/2} \, 1i_{13/2}{}^{-1}$
                                  & 13.21 & $\pi\, 1i_{13/2} \, 1h_{11/2}{}^{-1} \otimes 13^-_{  1}$
                                                    & 12.34 & $12^-_{  1} \otimes 13^-_{  1}$ \\
$26^-_1$ &13.5360& 13.67  & $\nu\, 1j_{15/2} \, 1i_{13/2}{}^{-1}
                             \otimes \nu\, 1i_{11/2} \, 1i_{13/2}{}^{-1}$
                                  & 13.04 & $\nu\, 1i_{11/2} \, 1i_{13/2}{}^{-1} \otimes 14^-_{  1}$
                                                    & 12.92 & $14^-_{  1} \otimes 12^+_{  1}$ \\
$26^+_1$ &       & 14.13 & $\nu\, 1j_{15/2} \, 1i_{13/2}{}^{-1}
                            \otimes \pi\, 1i_{13/2} \, 1h_{11/2}{}^{-1}$
                                  & 13.27 & $\nu\, 1j_{15/2} \, 1i_{13/2}{}^{-1} \otimes 12^-_{  1}$
                                                    & 12.38 & $13^-_{  1} \otimes 13^-_{  1}$ \\
$27^+_1$ &       & 14.23 & $\nu\, 1j_{15/2} \, 1i_{13/2}{}^{-1}
                            \otimes \nu\, 1j_{15/2} \, 1i_{13/2}{}^{-1}$
                                  & 13.31 & $\nu\, 1j_{15/2} \, 1i_{13/2}{}^{-1} \otimes 13^-_{  1}$
                                                    & 12.68 & $14^-_{  1} \otimes 13^-_{  1}$ \\
\end{tabular}
\end{ruledtabular}
\footnotetext[1]{The spin-parity assignment to the level 9.061
is $(17^+)$ in the NDS~\cite{NDS_2007} and $17^+$ in Ref.~\cite{Broda_2017} but
$15^+$ in Ref.~\cite{Heusler_2020}.}
\footnotetext[2]{The level 8.562 is taken from Ref.~\cite{NDS_2007}}
\end{table*}


Table \ref{table:multiphonon} also elucidates the generally rotational
trend $E_I \propto I(I+1)$.  The trend is the same for all
levels of approach such that we can learn about the underlying
structure already from the simple $N$p$N$h configurations.  The trend is
driven by a change in the angular momentum of one nucleon after the
other.  The spins of the high-spin yrast states are composed of
single-particle states having large angular momentum.  The
rotational part of a s.p. kinetic energy gives a large contribution to
the s.p. energy. Thus a change in the angular momentum of one nucleon
produces a significant change in the excitation energy. This rotation
is not collective but is a single-particle rotation. At first glance,
it resembles the rotational alignment known from rotating deformed
nuclei. However, the situation is different. Rotational alignment
dissolves gradually the collective cranking rotations which then are
replaced by a sequence of single-particle contributions. The doubly magic
$^{208}$Pb has a large spectral gap which inhibits cranking from the
onset. The generation of rotational spectra by single-particle
structures starts right away without a phase of collective rotations
coming before.

\bibliographystyle{apsrev4-1}
\bibliography{TTT}

\begin{thebibliography}{31}%
\makeatletter
\providecommand \@ifxundefined [1]{%
 \@ifx{#1\undefined}
}%
\providecommand \@ifnum [1]{%
 \ifnum #1\expandafter \@firstoftwo
 \else \expandafter \@secondoftwo
 \fi
}%
\providecommand \@ifx [1]{%
 \ifx #1\expandafter \@firstoftwo
 \else \expandafter \@secondoftwo
 \fi
}%
\providecommand \natexlab [1]{#1}%
\providecommand \enquote  [1]{``#1''}%
\providecommand \bibnamefont  [1]{#1}%
\providecommand \bibfnamefont [1]{#1}%
\providecommand \citenamefont [1]{#1}%
\providecommand \href@noop [0]{\@secondoftwo}%
\providecommand \href [0]{\begingroup \@sanitize@url \@href}%
\providecommand \@href[1]{\@@startlink{#1}\@@href}%
\providecommand \@@href[1]{\endgroup#1\@@endlink}%
\providecommand \@sanitize@url [0]{\catcode `\\12\catcode `\$12\catcode
  `\&12\catcode `\#12\catcode `\^12\catcode `\_12\catcode `\%12\relax}%
\providecommand \@@startlink[1]{}%
\providecommand \@@endlink[0]{}%
\providecommand \url  [0]{\begingroup\@sanitize@url \@url }%
\providecommand \@url [1]{\endgroup\@href {#1}{\urlprefix }}%
\providecommand \urlprefix  [0]{URL }%
\providecommand \Eprint [0]{\href }%
\providecommand \doibase [0]{http://dx.doi.org/}%
\providecommand \selectlanguage [0]{\@gobble}%
\providecommand \bibinfo  [0]{\@secondoftwo}%
\providecommand \bibfield  [0]{\@secondoftwo}%
\providecommand \translation [1]{[#1]}%
\providecommand \BibitemOpen [0]{}%
\providecommand \bibitemStop [0]{}%
\providecommand \bibitemNoStop [0]{.\EOS\space}%
\providecommand \EOS [0]{\spacefactor3000\relax}%
\providecommand \BibitemShut  [1]{\csname bibitem#1\endcsname}%
\let\auto@bib@innerbib\@empty
\bibitem [{\citenamefont {de~Voigt}\ \emph {et~al.}(1983)\citenamefont
  {de~Voigt}, \citenamefont {Dudek},\ and\ \citenamefont
  {Szyma\ifmmode~\acute{n}\else \'{n}\fi{}ski}}]{VoigtRMP83}%
  \BibitemOpen
  \bibfield  {author} {\bibinfo {author} {\bibfnamefont {M.~J.~A.}\
  \bibnamefont {de~Voigt}}, \bibinfo {author} {\bibfnamefont {J.}~\bibnamefont
  {Dudek}}, \ and\ \bibinfo {author} {\bibfnamefont {Z.}~\bibnamefont
  {Szyma\ifmmode~\acute{n}\else \'{n}\fi{}ski}},\ }\href {\doibase
  10.1103/RevModPhys.55.949} {\bibfield  {journal} {\bibinfo  {journal} {Rev.
  Mod. Phys.}\ }\textbf {\bibinfo {volume} {55}},\ \bibinfo {pages} {949}
  (\bibinfo {year} {1983})}\BibitemShut {NoStop}%
\bibitem [{\citenamefont {Saladin}\ \emph {et~al.}(1991)\citenamefont
  {Saladin}, \citenamefont {Sorensen},\ and\ \citenamefont
  {Vincent}}]{SaladinBook91}%
  \BibitemOpen
  \bibfield  {author} {\bibinfo {author} {\bibfnamefont {J.~X.}\ \bibnamefont
  {Saladin}}, \bibinfo {author} {\bibfnamefont {R.~A.}\ \bibnamefont
  {Sorensen}}, \ and\ \bibinfo {author} {\bibfnamefont {C.~M.}\ \bibnamefont
  {Vincent}},\ }\href {\doibase 10.1142/1242} {\emph {\bibinfo {title} {High
  Spin Physics and Gamma-Soft Nuclei}}}\ (\bibinfo  {publisher} {World
  Scientific},\ \bibinfo {year} {1991})\BibitemShut {NoStop}%
\bibitem [{\citenamefont {Ward}\ and\ \citenamefont {Fallon}(2001)}]{Ward2001}%
  \BibitemOpen
  \bibfield  {author} {\bibinfo {author} {\bibfnamefont {D.}~\bibnamefont
  {Ward}}\ and\ \bibinfo {author} {\bibfnamefont {P.}~\bibnamefont {Fallon}},\
  }\enquote {\bibinfo {title} {High spin properties of atomic nuclei},}\ in\
  \href {\doibase 10.1007/0-306-47915-X_3} {\emph {\bibinfo {booktitle}
  {Advances in Nuclear Physics}}},\ \bibinfo {editor} {edited by\ \bibinfo
  {editor} {\bibfnamefont {J.~W.}\ \bibnamefont {Negele}}\ and\ \bibinfo
  {editor} {\bibfnamefont {E.~W.}\ \bibnamefont {Vogt}}}\ (\bibinfo
  {publisher} {Springer US},\ \bibinfo {address} {Boston, MA},\ \bibinfo {year}
  {2001})\ pp.\ \bibinfo {pages} {167--291}\BibitemShut {NoStop}%
\bibitem [{\citenamefont {Broda}\ \emph {et~al.}(2017)\citenamefont {Broda},
  \citenamefont {Janssens}, \citenamefont {Iskra}, \citenamefont {Wrzesinski},
  \citenamefont {Fornal}, \citenamefont {Carpenter}, \citenamefont {Chiara},
  \citenamefont {Cieplicka-Ory\ifmmode~\acute{n}\else \'{n}\fi{}czak},
  \citenamefont {Hoffman}, \citenamefont {Kondev}, \citenamefont {Kr\'olas},
  \citenamefont {Lauritsen}, \citenamefont {Podolyak}, \citenamefont
  {Seweryniak}, \citenamefont {Shand}, \citenamefont {Szpak}, \citenamefont
  {Walters}, \citenamefont {Zhu},\ and\ \citenamefont {Brown}}]{Broda_2017}%
  \BibitemOpen
  \bibfield  {author} {\bibinfo {author} {\bibfnamefont {R.}~\bibnamefont
  {Broda}}, \bibinfo {author} {\bibfnamefont {R.~V.~F.}\ \bibnamefont
  {Janssens}}, \bibinfo {author} {\bibfnamefont {L.~W.}\ \bibnamefont {Iskra}},
  \bibinfo {author} {\bibfnamefont {J.}~\bibnamefont {Wrzesinski}}, \bibinfo
  {author} {\bibfnamefont {B.}~\bibnamefont {Fornal}}, \bibinfo {author}
  {\bibfnamefont {M.~P.}\ \bibnamefont {Carpenter}}, \bibinfo {author}
  {\bibfnamefont {C.~J.}\ \bibnamefont {Chiara}}, \bibinfo {author}
  {\bibfnamefont {N.}~\bibnamefont {Cieplicka-Ory\ifmmode~\acute{n}\else
  \'{n}\fi{}czak}}, \bibinfo {author} {\bibfnamefont {C.~R.}\ \bibnamefont
  {Hoffman}}, \bibinfo {author} {\bibfnamefont {F.~G.}\ \bibnamefont {Kondev}},
  \bibinfo {author} {\bibfnamefont {W.}~\bibnamefont {Kr\'olas}}, \bibinfo
  {author} {\bibfnamefont {T.}~\bibnamefont {Lauritsen}}, \bibinfo {author}
  {\bibfnamefont {Z.}~\bibnamefont {Podolyak}}, \bibinfo {author}
  {\bibfnamefont {D.}~\bibnamefont {Seweryniak}}, \bibinfo {author}
  {\bibfnamefont {C.~M.}\ \bibnamefont {Shand}}, \bibinfo {author}
  {\bibfnamefont {B.}~\bibnamefont {Szpak}}, \bibinfo {author} {\bibfnamefont
  {W.~B.}\ \bibnamefont {Walters}}, \bibinfo {author} {\bibfnamefont
  {S.}~\bibnamefont {Zhu}}, \ and\ \bibinfo {author} {\bibfnamefont {B.~A.}\
  \bibnamefont {Brown}},\ }\href {\doibase 10.1103/PhysRevC.95.064308}
  {\bibfield  {journal} {\bibinfo  {journal} {Phys. Rev. C}\ }\textbf {\bibinfo
  {volume} {95}},\ \bibinfo {pages} {064308} (\bibinfo {year}
  {2017})}\BibitemShut {NoStop}%
\bibitem [{\citenamefont {Bender}\ \emph {et~al.}(2003)\citenamefont {Bender},
  \citenamefont {Heenen},\ and\ \citenamefont {Reinhard}}]{Bender_2003}%
  \BibitemOpen
  \bibfield  {author} {\bibinfo {author} {\bibfnamefont {M.}~\bibnamefont
  {Bender}}, \bibinfo {author} {\bibfnamefont {P.-H.}\ \bibnamefont {Heenen}},
  \ and\ \bibinfo {author} {\bibfnamefont {P.-G.}\ \bibnamefont {Reinhard}},\
  }\href {\doibase 10.1103/Rev.Mod.Phys.75.121} {\bibfield  {journal} {\bibinfo
   {journal} {Rev. Mod. Phys.}\ }\textbf {\bibinfo {volume} {75}},\ \bibinfo
  {pages} {121} (\bibinfo {year} {2003})}\BibitemShut {NoStop}%
\bibitem [{\citenamefont {Vretenar}\ \emph {et~al.}(2005)\citenamefont
  {Vretenar}, \citenamefont {Afanasjev}, \citenamefont {Lalazissis},\ and\
  \citenamefont {Ring}}]{Vretenar_2005}%
  \BibitemOpen
  \bibfield  {author} {\bibinfo {author} {\bibfnamefont {D.}~\bibnamefont
  {Vretenar}}, \bibinfo {author} {\bibfnamefont {A.~V.}\ \bibnamefont
  {Afanasjev}}, \bibinfo {author} {\bibfnamefont {G.}~\bibnamefont
  {Lalazissis}}, \ and\ \bibinfo {author} {\bibfnamefont {P.}~\bibnamefont
  {Ring}},\ }\href {\doibase 10.1016/j.physrep.2004.10.001} {\bibfield
  {journal} {\bibinfo  {journal} {Phys. Rep.}\ }\textbf {\bibinfo {volume}
  {409}},\ \bibinfo {pages} {101} (\bibinfo {year} {2005})}\BibitemShut
  {NoStop}%
\bibitem [{\citenamefont {Erler}\ \emph {et~al.}(2011)\citenamefont {Erler},
  \citenamefont {Kl{\"{u}}pfel},\ and\ \citenamefont {Reinhard}}]{Erler_2011}%
  \BibitemOpen
  \bibfield  {author} {\bibinfo {author} {\bibfnamefont {J.}~\bibnamefont
  {Erler}}, \bibinfo {author} {\bibfnamefont {P.}~\bibnamefont
  {Kl{\"{u}}pfel}}, \ and\ \bibinfo {author} {\bibfnamefont {P.~G.}\
  \bibnamefont {Reinhard}},\ }\href {doi:10.1088/0954-3899/38/3/033101}
  {\bibfield  {journal} {\bibinfo  {journal} {J. Phys. G}\ }\textbf {\bibinfo
  {volume} {38}},\ \bibinfo {pages} {033101} (\bibinfo {year}
  {2011})}\BibitemShut {NoStop}%
\bibitem [{\citenamefont {Dro\.zd\.z}\ \emph {et~al.}(1990)\citenamefont
  {Dro\.zd\.z}, \citenamefont {Nishizaki}, \citenamefont {Speth},\ and\
  \citenamefont {Wambach}}]{Drozdz_1990}%
  \BibitemOpen
  \bibfield  {author} {\bibinfo {author} {\bibfnamefont {S.}~\bibnamefont
  {Dro\.zd\.z}}, \bibinfo {author} {\bibfnamefont {S.}~\bibnamefont
  {Nishizaki}}, \bibinfo {author} {\bibfnamefont {J.}~\bibnamefont {Speth}}, \
  and\ \bibinfo {author} {\bibfnamefont {J.}~\bibnamefont {Wambach}},\ }\href
  {\doibase https://doi.org/10.1016/0370-1573(90)90084-F} {\bibfield  {journal}
  {\bibinfo  {journal} {Phys. Rep.}\ }\textbf {\bibinfo {volume} {197}},\
  \bibinfo {pages} {1} (\bibinfo {year} {1990})}\BibitemShut {NoStop}%
\bibitem [{\citenamefont {Soloviev}(1992)}]{Soloviev_1992}%
  \BibitemOpen
  \bibfield  {author} {\bibinfo {author} {\bibfnamefont {V.~G.}\ \bibnamefont
  {Soloviev}},\ }\href@noop {} {\emph {\bibinfo {title} {Theory of Atomic
  Nuclei: Quasiparticles and Phonons}}}\ (\bibinfo  {publisher} {Institute of
  Physics},\ \bibinfo {address} {Bristol and Philadelphia},\ \bibinfo {year}
  {1992})\BibitemShut {NoStop}%
\bibitem [{\citenamefont {Col\`o}\ and\ \citenamefont
  {Bortignon}(2001)}]{Col01a}%
  \BibitemOpen
  \bibfield  {author} {\bibinfo {author} {\bibfnamefont {G.}~\bibnamefont
  {Col\`o}}\ and\ \bibinfo {author} {\bibfnamefont {P.~F.}\ \bibnamefont
  {Bortignon}},\ }\href {\doibase
  https://doi.org/10.1016/S0375-9474(01)00633-9} {\bibfield  {journal}
  {\bibinfo  {journal} {Nucl. Phys. A}\ }\textbf {\bibinfo {volume} {687}},\
  \bibinfo {pages} {282c} (\bibinfo {year} {2001})}\BibitemShut {NoStop}%
\bibitem [{\citenamefont {Kamerdzhiev}\ \emph {et~al.}(2004)\citenamefont
  {Kamerdzhiev}, \citenamefont {Speth},\ and\ \citenamefont
  {Tertychny}}]{Kamerdzhiev_2004}%
  \BibitemOpen
  \bibfield  {author} {\bibinfo {author} {\bibfnamefont {S.}~\bibnamefont
  {Kamerdzhiev}}, \bibinfo {author} {\bibfnamefont {J.}~\bibnamefont {Speth}},
  \ and\ \bibinfo {author} {\bibfnamefont {G.}~\bibnamefont {Tertychny}},\
  }\href {\doibase 10.1016/j.physrep.2003.11.001} {\bibfield  {journal}
  {\bibinfo  {journal} {Phys. Rep.}\ }\textbf {\bibinfo {volume} {393}},\
  \bibinfo {pages} {1} (\bibinfo {year} {2004})}\BibitemShut {NoStop}%
\bibitem [{\citenamefont {Lyutorovich}\ \emph {et~al.}(2012)\citenamefont
  {Lyutorovich}, \citenamefont {Tselyaev}, \citenamefont {Speth}, \citenamefont
  {Krewald}, \citenamefont {Gr{\"{u}}mmer},\ and\ \citenamefont
  {Reinhard}}]{Lyutorovich_2012}%
  \BibitemOpen
  \bibfield  {author} {\bibinfo {author} {\bibfnamefont {N.}~\bibnamefont
  {Lyutorovich}}, \bibinfo {author} {\bibfnamefont {V.~I.}\ \bibnamefont
  {Tselyaev}}, \bibinfo {author} {\bibfnamefont {J.}~\bibnamefont {Speth}},
  \bibinfo {author} {\bibfnamefont {S.}~\bibnamefont {Krewald}}, \bibinfo
  {author} {\bibfnamefont {F.}~\bibnamefont {Gr{\"{u}}mmer}}, \ and\ \bibinfo
  {author} {\bibfnamefont {P.~G.}\ \bibnamefont {Reinhard}},\ }\href {\doibase
  10.1103/PhysRevLett.109.092502} {\bibfield  {journal} {\bibinfo  {journal}
  {Phys. Rev. Lett.}\ }\textbf {\bibinfo {volume} {109}},\ \bibinfo {pages}
  {092502} (\bibinfo {year} {2012})}\BibitemShut {NoStop}%
\bibitem [{\citenamefont {Kl{\"{u}}pfel}\ \emph {et~al.}(2009)\citenamefont
  {Kl{\"{u}}pfel}, \citenamefont {Reinhard}, \citenamefont {B{\"{u}}rvenich},\
  and\ \citenamefont {Maruhn}}]{Kluepfel_2009}%
  \BibitemOpen
  \bibfield  {author} {\bibinfo {author} {\bibfnamefont {P.}~\bibnamefont
  {Kl{\"{u}}pfel}}, \bibinfo {author} {\bibfnamefont {P.~G.}\ \bibnamefont
  {Reinhard}}, \bibinfo {author} {\bibfnamefont {T.~J.}\ \bibnamefont
  {B{\"{u}}rvenich}}, \ and\ \bibinfo {author} {\bibfnamefont {J.~A.}\
  \bibnamefont {Maruhn}},\ }\href {\doibase 10.1103/PhysRevC.79.034310}
  {\bibfield  {journal} {\bibinfo  {journal} {Phys. Rev. C}\ }\textbf {\bibinfo
  {volume} {79}},\ \bibinfo {pages} {034310} (\bibinfo {year}
  {2009})}\BibitemShut {NoStop}%
\bibitem [{\citenamefont {Bertulani}\ and\ \citenamefont
  {Ponomarev}(1999)}]{Bertulani_1999}%
  \BibitemOpen
  \bibfield  {author} {\bibinfo {author} {\bibfnamefont {C.~A.}\ \bibnamefont
  {Bertulani}}\ and\ \bibinfo {author} {\bibfnamefont {V.~Y.}\ \bibnamefont
  {Ponomarev}},\ }\href {\doibase
  https://doi.org/10.1016/S0370-1573(99)00038-1} {\bibfield  {journal}
  {\bibinfo  {journal} {Phys. Rep.}\ }\textbf {\bibinfo {volume} {321}},\
  \bibinfo {pages} {139} (\bibinfo {year} {1999})}\BibitemShut {NoStop}%
\bibitem [{\citenamefont {Brown}(2000)}]{Brown_2000}%
  \BibitemOpen
  \bibfield  {author} {\bibinfo {author} {\bibfnamefont {B.~A.}\ \bibnamefont
  {Brown}},\ }\href {\doibase 10.1103/PhysRevLett.85.5300} {\bibfield
  {journal} {\bibinfo  {journal} {Phys. Rev. Lett.}\ }\textbf {\bibinfo
  {volume} {85}},\ \bibinfo {pages} {5300} (\bibinfo {year}
  {2000})}\BibitemShut {NoStop}%
\bibitem [{\citenamefont {Litvinova}\ \emph {et~al.}(2010)\citenamefont
  {Litvinova}, \citenamefont {Ring},\ and\ \citenamefont
  {Tselyaev}}]{Litvinova_2010}%
  \BibitemOpen
  \bibfield  {author} {\bibinfo {author} {\bibfnamefont {E.}~\bibnamefont
  {Litvinova}}, \bibinfo {author} {\bibfnamefont {P.}~\bibnamefont {Ring}}, \
  and\ \bibinfo {author} {\bibfnamefont {V.}~\bibnamefont {Tselyaev}},\ }\href
  {\doibase 10.1103/PhysRevLett.105.022502} {\bibfield  {journal} {\bibinfo
  {journal} {Phys. Rev. Lett.}\ }\textbf {\bibinfo {volume} {105}},\ \bibinfo
  {pages} {022502} (\bibinfo {year} {2010})}\BibitemShut {NoStop}%
\bibitem [{\citenamefont {Litvinova}\ \emph {et~al.}(2013)\citenamefont
  {Litvinova}, \citenamefont {Ring},\ and\ \citenamefont
  {Tselyaev}}]{Litvinova_2013}%
  \BibitemOpen
  \bibfield  {author} {\bibinfo {author} {\bibfnamefont {E.}~\bibnamefont
  {Litvinova}}, \bibinfo {author} {\bibfnamefont {P.}~\bibnamefont {Ring}}, \
  and\ \bibinfo {author} {\bibfnamefont {V.}~\bibnamefont {Tselyaev}},\ }\href
  {\doibase https://doi.org/10.1103/PhysRevC.88.044320} {\bibfield  {journal}
  {\bibinfo  {journal} {Phys. Rev. C}\ }\textbf {\bibinfo {volume} {88}},\
  \bibinfo {pages} {044320} (\bibinfo {year} {2013})}\BibitemShut {NoStop}%
\bibitem [{\citenamefont {Tselyaev}\ \emph {et~al.}(2018)\citenamefont
  {Tselyaev}, \citenamefont {Lyutorovich}, \citenamefont {Speth},\ and\
  \citenamefont {Reinhard}}]{Tselyaev_2018}%
  \BibitemOpen
  \bibfield  {author} {\bibinfo {author} {\bibfnamefont {V.}~\bibnamefont
  {Tselyaev}}, \bibinfo {author} {\bibfnamefont {N.}~\bibnamefont
  {Lyutorovich}}, \bibinfo {author} {\bibfnamefont {J.}~\bibnamefont {Speth}},
  \ and\ \bibinfo {author} {\bibfnamefont {P.-G.}\ \bibnamefont {Reinhard}},\
  }\href {\doibase 10.1103/PhysRevC.97.044308} {\bibfield  {journal} {\bibinfo
  {journal} {Phys. Rev. C}\ }\textbf {\bibinfo {volume} {97}},\ \bibinfo
  {pages} {044308} (\bibinfo {year} {2018})}\BibitemShut {NoStop}%
\bibitem [{\citenamefont {Tselyaev}(1989)}]{Tselyaev_1989}%
  \BibitemOpen
  \bibfield  {author} {\bibinfo {author} {\bibfnamefont {V.~I.}\ \bibnamefont
  {Tselyaev}},\ }\href@noop {} {\bibfield  {journal} {\bibinfo  {journal} {Sov.
  J. Nucl. Phys.}\ }\textbf {\bibinfo {volume} {50}},\ \bibinfo {pages} {780}
  (\bibinfo {year} {1989})}\BibitemShut {NoStop}%
\bibitem [{\citenamefont {Tselyaev}(2007)}]{Tselyaev_2007}%
  \BibitemOpen
  \bibfield  {author} {\bibinfo {author} {\bibfnamefont {V.~I.}\ \bibnamefont
  {Tselyaev}},\ }\href {\doibase 10.1103/Phys.RevC.75.024306} {\bibfield
  {journal} {\bibinfo  {journal} {Phys. Rev. C}\ }\textbf {\bibinfo {volume}
  {75}},\ \bibinfo {pages} {024306} (\bibinfo {year} {2007})}\BibitemShut
  {NoStop}%
\bibitem [{\citenamefont {Lyutorovich}\ \emph {et~al.}(2015)\citenamefont
  {Lyutorovich}, \citenamefont {Tselyaev}, \citenamefont {Speth}, \citenamefont
  {Krewald}, \citenamefont {Gr\"ummer},\ and\ \citenamefont
  {Reinhard}}]{Lyutorovich_2015}%
  \BibitemOpen
  \bibfield  {author} {\bibinfo {author} {\bibfnamefont {N.}~\bibnamefont
  {Lyutorovich}}, \bibinfo {author} {\bibfnamefont {V.}~\bibnamefont
  {Tselyaev}}, \bibinfo {author} {\bibfnamefont {J.}~\bibnamefont {Speth}},
  \bibinfo {author} {\bibfnamefont {S.}~\bibnamefont {Krewald}}, \bibinfo
  {author} {\bibfnamefont {F.}~\bibnamefont {Gr\"ummer}}, \ and\ \bibinfo
  {author} {\bibfnamefont {P.-G.}\ \bibnamefont {Reinhard}},\ }\href {\doibase
  https://doi.org/10.1016/j.physletb.2015.08.003} {\bibfield  {journal}
  {\bibinfo  {journal} {Phys. Lett. B}\ }\textbf {\bibinfo {volume} {749}},\
  \bibinfo {pages} {292} (\bibinfo {year} {2015})}\BibitemShut {NoStop}%
\bibitem [{\citenamefont {Lyutorovich}\ \emph {et~al.}(2016)\citenamefont
  {Lyutorovich}, \citenamefont {Tselyaev}, \citenamefont {Speth}, \citenamefont
  {Krewald},\ and\ \citenamefont {Reinhard}}]{Lyutorovich_2016}%
  \BibitemOpen
  \bibfield  {author} {\bibinfo {author} {\bibfnamefont {N.}~\bibnamefont
  {Lyutorovich}}, \bibinfo {author} {\bibfnamefont {V.}~\bibnamefont
  {Tselyaev}}, \bibinfo {author} {\bibfnamefont {J.}~\bibnamefont {Speth}},
  \bibinfo {author} {\bibfnamefont {S.}~\bibnamefont {Krewald}}, \ and\
  \bibinfo {author} {\bibfnamefont {P.-G.}\ \bibnamefont {Reinhard}},\ }\href
  {ArXiv: http://arxiv.org/abs/1602.00862} {\bibfield  {journal} {\bibinfo
  {journal} {Phys. At. Nucl.}\ }\textbf {\bibinfo {volume} {79}},\ \bibinfo
  {pages} {868} (\bibinfo {year} {2016})}\BibitemShut {NoStop}%
\bibitem [{\citenamefont {Tselyaev}\ \emph {et~al.}(2016)\citenamefont
  {Tselyaev}, \citenamefont {Lyutorovich}, \citenamefont {Speth}, \citenamefont
  {Krewald},\ and\ \citenamefont {Reinhard}}]{Tselyaev_2016}%
  \BibitemOpen
  \bibfield  {author} {\bibinfo {author} {\bibfnamefont {V.}~\bibnamefont
  {Tselyaev}}, \bibinfo {author} {\bibfnamefont {N.}~\bibnamefont
  {Lyutorovich}}, \bibinfo {author} {\bibfnamefont {J.}~\bibnamefont {Speth}},
  \bibinfo {author} {\bibfnamefont {S.}~\bibnamefont {Krewald}}, \ and\
  \bibinfo {author} {\bibfnamefont {P.-G.}\ \bibnamefont {Reinhard}},\ }\href
  {\doibase 10.1103/PhysRevC.94.034306} {\bibfield  {journal} {\bibinfo
  {journal} {Phys. Rev. C}\ }\textbf {\bibinfo {volume} {94}},\ \bibinfo
  {pages} {034306} (\bibinfo {year} {2016})}\BibitemShut {NoStop}%
\bibitem [{\citenamefont {Tselyaev}\ \emph {et~al.}(2020)\citenamefont
  {Tselyaev}, \citenamefont {Lyutorovich}, \citenamefont {Speth},\ and\
  \citenamefont {Reinhard}}]{Tselyaev_2020}%
  \BibitemOpen
  \bibfield  {author} {\bibinfo {author} {\bibfnamefont {V.}~\bibnamefont
  {Tselyaev}}, \bibinfo {author} {\bibfnamefont {N.}~\bibnamefont
  {Lyutorovich}}, \bibinfo {author} {\bibfnamefont {J.}~\bibnamefont {Speth}},
  \ and\ \bibinfo {author} {\bibfnamefont {P.-G.}\ \bibnamefont {Reinhard}},\
  }\href {\doibase 10.1103/PhysRevC.102.064319} {\bibfield  {journal} {\bibinfo
   {journal} {Phys. Rev. C}\ }\textbf {\bibinfo {volume} {102}},\ \bibinfo
  {pages} {064319} (\bibinfo {year} {2020})}\BibitemShut {NoStop}%
\bibitem [{\citenamefont {Kl{\"{u}}pfel}\ \emph {et~al.}(2008)\citenamefont
  {Kl{\"{u}}pfel}, \citenamefont {Erler}, \citenamefont {Reinhard},\ and\
  \citenamefont {Maruhn}}]{Kluepfel_2008}%
  \BibitemOpen
  \bibfield  {author} {\bibinfo {author} {\bibfnamefont {P.}~\bibnamefont
  {Kl{\"{u}}pfel}}, \bibinfo {author} {\bibfnamefont {J.}~\bibnamefont
  {Erler}}, \bibinfo {author} {\bibfnamefont {P.-G.}\ \bibnamefont {Reinhard}},
  \ and\ \bibinfo {author} {\bibfnamefont {J.~A.}\ \bibnamefont {Maruhn}},\
  }\href {http://dx.doi.org/10.1140/epja/i2008-10633-3} {\bibfield  {journal}
  {\bibinfo  {journal} {Eur. Phys. J. A}\ }\textbf {\bibinfo {volume} {37}},\
  \bibinfo {pages} {343} (\bibinfo {year} {2008})}\BibitemShut {NoStop}%
\bibitem [{\citenamefont {Pototzky}\ \emph {et~al.}(2010)\citenamefont
  {Pototzky}, \citenamefont {Erler}, \citenamefont {Reinhard},\ and\
  \citenamefont {Nesterenko}}]{Pot10a}%
  \BibitemOpen
  \bibfield  {author} {\bibinfo {author} {\bibfnamefont {K.~J.}\ \bibnamefont
  {Pototzky}}, \bibinfo {author} {\bibfnamefont {J.}~\bibnamefont {Erler}},
  \bibinfo {author} {\bibfnamefont {P.-G.}\ \bibnamefont {Reinhard}}, \ and\
  \bibinfo {author} {\bibfnamefont {V.~O.}\ \bibnamefont {Nesterenko}},\ }\href
  {http://dx.doi.org/10.1140/epja/i2010-11045-6} {\bibfield  {journal}
  {\bibinfo  {journal} {Eur. Phys. J. A}\ }\textbf {\bibinfo {volume} {46}},\
  \bibinfo {pages} {299} (\bibinfo {year} {2010})}\BibitemShut {NoStop}%
\bibitem [{\citenamefont {Tselyaev}\ \emph {et~al.}(2019)\citenamefont
  {Tselyaev}, \citenamefont {Lyutorovich}, \citenamefont {Speth}, \citenamefont
  {Reinhard},\ and\ \citenamefont {Smirnov}}]{Tselyaev_2019}%
  \BibitemOpen
  \bibfield  {author} {\bibinfo {author} {\bibfnamefont {V.}~\bibnamefont
  {Tselyaev}}, \bibinfo {author} {\bibfnamefont {N.}~\bibnamefont
  {Lyutorovich}}, \bibinfo {author} {\bibfnamefont {J.}~\bibnamefont {Speth}},
  \bibinfo {author} {\bibfnamefont {P.-G.}\ \bibnamefont {Reinhard}}, \ and\
  \bibinfo {author} {\bibfnamefont {D.}~\bibnamefont {Smirnov}},\ }\href
  {\doibase 10.1103/PhysRevC.99.064329} {\bibfield  {journal} {\bibinfo
  {journal} {Phys. Rev. C}\ }\textbf {\bibinfo {volume} {99}},\ \bibinfo
  {pages} {064329} (\bibinfo {year} {2019})}\BibitemShut {NoStop}%
\bibitem [{\citenamefont {Brown}(1998)}]{Brown_1998}%
  \BibitemOpen
  \bibfield  {author} {\bibinfo {author} {\bibfnamefont {B.~A.}\ \bibnamefont
  {Brown}},\ }\href@noop {} {\bibfield  {journal} {\bibinfo  {journal} {Phys.
  Rev. C}\ }\textbf {\bibinfo {volume} {58}},\ \bibinfo {pages} {220} (\bibinfo
  {year} {1998})}\BibitemShut {NoStop}%
\bibitem [{\citenamefont {Martin}(2007)}]{NDS_2007}%
  \BibitemOpen
  \bibfield  {author} {\bibinfo {author} {\bibfnamefont {M.}~\bibnamefont
  {Martin}},\ }\href {\doibase http://dx.doi.org/10.1016/j.nds.2007.07.001}
  {\bibfield  {journal} {\bibinfo  {journal} {Nuclear Data Sheets}\ }\textbf
  {\bibinfo {volume} {108}},\ \bibinfo {pages} {1583 } (\bibinfo {year}
  {2007})}\BibitemShut {NoStop}%
\bibitem [{\citenamefont {Ring}\ and\ \citenamefont {Schuck}(1980)}]{Rin80aB}%
  \BibitemOpen
  \bibfield  {author} {\bibinfo {author} {\bibfnamefont {P.}~\bibnamefont
  {Ring}}\ and\ \bibinfo {author} {\bibfnamefont {P.}~\bibnamefont {Schuck}},\
  }\href@noop {} {\emph {\bibinfo {title} {The Nuclear Many-Body Problem}}}\
  (\bibinfo  {publisher} {Springer--Verl.},\ \bibinfo {address} {New York,
  Heidelberg, Berlin},\ \bibinfo {year} {1980})\BibitemShut {NoStop}%
\bibitem [{\citenamefont {Heusler}(2020)}]{Heusler_2020}%
  \BibitemOpen
  \bibfield  {author} {\bibinfo {author} {\bibfnamefont {A.}~\bibnamefont
  {Heusler}},\ }\href {\doibase 10.1088/1742-6596/1643/1/012137} {\bibfield
  {journal} {\bibinfo  {journal} {Journal of Physics: Conference Series}\
  }\textbf {\bibinfo {volume} {1643}},\ \bibinfo {pages} {012137} (\bibinfo
  {year} {2020})}\BibitemShut {NoStop}%
\end{thebibliography}%

\end{document}